# Hybrid *ab initio* method for examining thermal properties in magnetic materials

Matthew Heine,[1] Olle Hellman[2] and David Broido[1*]

[1]Department of Physics, Boston College, Chestnut Hill, Massachusetts 02467, USA

[2]Department of Materials and Interfaces, Weizmann Institute of Science, Rehovot 76100, Israel

Abstract-

A hybrid *ab initio* theoretical approach for examining thermal properties in magnetic systems of unknown entropy is presented. Commonly used theoretical approaches interrogate thermal properties from Gibbs/Helmholtz free energies, which require an accurate model of magnetic interactions. The present approach avoids this requirement by instead calculating system pressure from thermally disordered microstates that properly incorporate vibrational and spin subsystems at each temperature as well as the coupling between these subsystems. In place of a specific model for magnetic interactions, the approach integrates measurements of temperature dependent magnetization of the studied material. We apply the approach to calculate phonon modes and to investigate the anomalously low thermal expansion of the classical Invar alloy, $Fe_{0.65}Ni_{0.35}$. The calculated phonon dispersions for Invar are in excellent agreement with measured data. The Invar thermal expansion is shown to remain small between 50 K and room temperature, consistent with the experimentally observed low thermal expansion value in this same temperature range. This anomalously small thermal expansion is directly connected to a small positive contribution from lattice thermal disorder that is nearly canceled by a large negative magnetic disorder contribution. By contrast, calculations for bcc Fe show a much larger thermal expansion, consistent with experiment, which is dominated by a large contribution from lattice thermal disorder that is reduced only slightly by a small negative contribution from that of magnetism. These findings give insights into the unusual nature of magnetism and spin-lattice coupling in Invar and Fe, and they support the presented new methodology as a complementary way to investigate thermal properties of magnetic materials.

* broido@bc.edu

## I. Introduction

A common theoretical approach to calculate a material's thermal properties starts from the Helmholtz Free energy, $F = F(V,T) = E - TS$, where $E$ is the energy, $T$ is temperature, $V$ is the crystal volume, and $S$ is the entropy. At each $T$, the equilibrium volume, $V(T)$, is the one that makes the pressure, $P = -(\partial F/\partial V)_T$, vanish. Identifying this volume requires calculations of temperature-dependent phonons. Once it is obtained, thermal expansion can be assessed from the change in $V$ with $T$ along the curve $V(T, P = 0)$ i.e. the one for which $P$ vanishes. Implementation of this theoretical approach requires an accurate model for $F$ and hence the entropy. In magnetic materials that have accepted theoretical models of magnetic interactions, phonon modes renormalized by both anharmonicity and magnetism as well as spin-lattice coupling can be obtained [1, 2], thus giving $S_{vib}$. Similarly, knowledge of magnetic interactions allows the construction of models for $S_{mag}$.

But what if an accepted model for the magnetism does not exist? In the present work, we introduce a complementary hybrid *ab initio* theoretical approach to calculate phonons and to examine thermal properties such as thermal expansion in a magnetic material. The theory bypasses the need to define the free energy, or more generally, the magnetic entropy. Instead, configurational averages over microstates of thermally disordered atomic displacements and magnetic moment orientations are connected to the measured temperature-dependent magnetization. Using this input from measurement, the approach extracts temperature dependent phonon modes renormalized by both anharmonicity and spin-lattice coupling. In addition, by identifying the $V(T)$ giving zero pressure at different $T$, the thermal expansion can be examined.

To test the hybrid *ab initio* theoretical approach, we apply it to calculate phonons and to examine thermal expansion in the "classical Invar" alloy, $Fe_{0.65}Ni_{0.35}$, and in bcc Fe. It is well-known that Invar possesses an anomalously low thermal expansion coefficient around and below room temperature, which is referred to as the Invar effect [3-6]. The explanation for the anomalous thermal expansion has been an open question in condensed matter physics for over a century [5, 7-11], and a complete theoretical description of this behavior is outside the scope of the present work. Nevertheless, we endeavor to assess the merits of the developed theory by (i) comparing calculated phonon dispersions to experiment and (ii) investigating the commonly held notion that the Invar effect arises from a near perfect cancellation between a positive contribution to thermal expansion by the vibrating lattice and a negative contribution from the magnetic subsystem [11].

Excellent agreement with the measured room temperature (RT) phonon dispersions in Invar, which, to our knowledge, has not been demonstrated before. Remarkably, the calculations also identify a near-zero thermal expansion for Invar in the range 50K to 300K, consistent with measurement. The findings suggest that the small thermal expansion in Invar results from a reduced contribution from thermal lattice disorder that is nearly canceled by a large negative contribution from thermal magnetic disorder. To add support of the developed theoretical approach, we also perform calculations for bcc Fe for which accepted models of the magnetic interactions exist. Excellent agreement with the measured RT phonon dispersions of bcc Fe is obtained using the new approach. In contrast to Invar, a much larger thermal expansion is found for Fe, in an equivalent temperature range, whose value is consistent with experiment. For Fe, the lattice thermal disorder contribution to thermal expansion dominates while the analogous contribution from the magnetic subsystem is found to be small.

## II. Thermal properties from configurational averaging of microstates

As noted above, a common theoretical approach to calculate phonons and thermal properties of a material such as thermal expansion involves determination of $V(T, P = 0)$ at each $T$, by first calculating the free energy, $F(V,T)$ and then determining where $P = -(\partial F/\partial V)_T$ vanishes. The Helmholtz free energy, $F = E - TS$, contains both vibrational and magnetic contributions. The present approach bypasses the need to explicitly define the entropy in the following way. First, we express $F$ as: $F = -k_B T ln(Z(V,T))$, where $Z(V,T)$ is the partition function:

$$Z(V,T) = \sum_i exp(-E_i/k_B T) \qquad (1)$$

where the sum over $i$ is a sum over all microstates, and $E_i$ is the energy of microstate $i$. Then,

$$P(V,T) = -\left(\frac{\partial F}{\partial V}\right)_T = \sum_i p_i \left(-\frac{\partial E_i}{\partial V}\right) = \langle (P_i) \rangle_{V,T} \qquad (2)$$

where $p_i = exp(-E_i/k_B T)/Z$ is the probability of occurrence of microstate $i$, and in the last term in Eq. 2 the angle brackets $< ... >_{V,T}$ represent a configurational average over microstates in the canonical ensemble for a given $V$ and $T$. Here, we have defined a microstate pressure: $P_i = -(\partial E_i/\partial V)$. We note that $P_i$ is not a thermodynamic quantity but merely a notational convenience.

An important feature of using Eq. 2, is that it avoids dealing explicitly with the entropy. Entropy is not neglected; instead, it is dealt with implicitly through the configurational averaging. Conceptually, it may help to recall that entropy is a function of the microstate probability distribution, $p_i$, [12], $S = -k_B \sum_i p_i ln(p_i)$. Therefore, a choice of microstate averaging is effectively a choice of entropy.

But haven't we traded one problem for another? How are we to determine the microstates appropriate to each $V, T$ and which ones should be included in the configurational averaging? We now discuss the computational approach to accomplish this task.

### III. Computational Approach

In order to calculate the pressure, $P(V,T)$, from Eq. 2, we will need to obtain $P_i$ values and perform the configurational averaging. Toward that end, our first step will be to obtain the phonon modes. Then, we discuss how to calculate the microstate energies and demonstrate how to organize these so as to extract $P(V,T)$. All calculations are implemented within the framework of Density Functional Theory (DFT), as implemented in VASP [13-16].

*Phonons:* To calculate phonon modes, we use the temperature dependent effective potential (TDEP) scheme [1, 2, 17]. In the TDEP approach, a set of supercell "snapshots", each possessing thermally relevant atomic displacements and magnetic moment orientations, is generated for a set of {*V*, *T*}. In the present work, this is achieved using a stochastic sampling approach [18-20]. Here we use a 32-atom supercell. In principle, the chemical disorder of this random alloy may be treated in the same fashion, but for computational efficiency, the technique of special quasi-random structures (SQS) is used [21] to populate the supercell with Fe and Ni atoms in the proper proportion, as has be done previously with TDEP [22]. An example of such snapshots is shown for Invar in Fig. 1. Note that all atoms are thermally displaced, the magnetic moments are thermally disordered, and some atoms are Fe while others are Ni, according to the SQS calculation. A DFT calculation is then performed for each of these thermally and chemically disordered supercell snapshots in order to interrogate the interatomic potential energy surface by generating force-displacement relationships for the atoms.

The TDEP Hamiltonian is

$$H_{TDEP} = U_0 + \sum_{i,\alpha} \frac{P_{i,\alpha}^2}{2M_i} + \frac{1}{2} \sum_{i,j,\alpha,\beta} \Phi_{i,j}^{\alpha,\beta}(V,T) u_{i,\alpha} u_{j,\beta} \qquad (3)$$

where $\Phi_{i,j}^{\alpha,\beta}(V,T)$ are effective IFCs which are explicit functions of both *V* and *T*. Any number of IFCs may be used in our framework. For each (*V-T*), the effective IFCs are defined to be those that best reproduce, in the least-squares sense, the ensemble of thermally-relevant DFT-calculated force-displacement relationships described above. Thus these effective IFCs best reproduce the thermally-sampled Born-Oppenheimer (BO) energy surface, contrasted with frequently-constructed IFCs which are derivatives of the BO energy surface with respect to a static ideal lattice.

In this way, effects of spin-lattice coupling are captured *implicitly*; the effective IFCs are those that best reproduce *all* the effects present in the DFT calculations, including spin-lattice coupling as well as lattice anharmonicity, chemical disorder, and local environment effects. The end output of the calculation is therefore a monatomic fcc lattice of an effective atom with renormalized IFCs and renormalized phonon modes are calculated by diagonalizing the dynamical matrix.

*Calculation of microstates:* For each spin-lattice snapshot, $i$, a DFT calculation gives us an energy, $E_{DFT}[i]$. However, since the atoms in such a snapshot are thermally displaced but still stationary, this calculation neglects the kinetic energy of the ions. In actuality, each microstate energy is given by: $E_i = K_i + E_{DFT}[i]$, where $K_i$ is the ion kinetic energy of microstate $i$. In principle, the kinetic energy contribution could be obtained in a molecular dynamics (MD) calculation, but such a calculation would omit important quantum mechanical behavior. In particular, at low temperatures, zero-point motion can dominate the amplitudes of the atomic displacements. The zero-point motion is captured within our snapshots approach. Also, the time evolution of the noncollinear magnetic moments within such an MD calculation is a nontrivial problem to solve.

The thermodynamic pressure can then be expressed as:

$$P(V,T) = P^K(V,T) + \langle P_{DFT}[i] \rangle_{V,T} \tag{4}$$

where the first term is the contribution to pressure from the ion kinetic energy and the second term is the remaining contribution extracted from the DFT calculations: $\langle P_{DFT}[i] \rangle_{V,T} = \langle -\partial E_{DFT}[i]/\partial V \rangle_{V,T}$. In the quasi-harmonic approximation, it is straightforward to show that [23]:

$$P^K(V,T) = \frac{1}{2} P^{vib}(V,T) \tag{5}$$

where $P^{vib}$ is the contribution to the pressure from the vibrational energy, $E^{vib}$. It can be obtained in a straightforward way once the relevant microstates are identified.

The configurational averaging over magnetic states could also be accomplished if a theory describing magnetism in the material of interest were known. For example, in Fe the magnetic system is often described theoretically using the Heisenberg model. We have found excellent agreement with the temperature dependent phonon frequencies using the Heisenberg model within the TDEP framework [1]. The present approach is developed for cases where an accepted theory of the magnetic interactions does not exist. The theory proceeds by integrating experimental measurements of the normalized magnetization as a function of temperature to guide the selection of microstates to be used in the configurational averaging. A central finding of the present work is that, for both bcc Fe and Invar, the calculated total pressure can be directly specified by the normalized magnetization. Moreover, we find that configurational averaging over magnetic states is unnecessary; using a single magnetic "snapshot" consistent with measurement is sufficient for calculating pressure.

Each magnetic snapshot may be characterized by a normalized magnetization, $\mathfrak{M}$, defined as the net magnetization per atom of the supercell divided by the average local moment size. The total pressure is taken to be the pressure calculated using a single magnetic snapshot of normalized magnetization, $\mathfrak{M}$:

$$P(V,T) \approx P_{\mathfrak{M}}(V,T) \qquad (6)$$

In the calculation of $P_{\mathfrak{M}}$, averaging over various thermally-displaced atomic configurations is still performed, but only a single magnetic configuration is used. This approximation was validated by demonstrating that pressures derived from considering only one magnetic state according to Eq. 6 were equivalent to the pressure calculated when averaging over magnetic states with similar $\mathfrak{M}$.

A detailed description of this computational scheme and its justification for Invar are presented in Appendix A.

## IV. Results

To our knowledge, an *ab initio* calculation of phonon dispersions in $Fe_{0.65}Ni_{0.35}$ and demonstrated agreement with this measured data does not exist. Figure 2 shows the calculated room-temperature (296K) phonon dispersions using the presented computational approach compared to measured data at that same temperature [24-26]. Calculations are for a lattice constant of $a$ = 3.5845Å, which was found to give nearly zero pressure at room temperature, and a normalized magnetization value of $\mathfrak{M}$ = 0.8, consistent with the measured lattice constant [27] and magnetization of the Invar alloy [28-31] at room temperature. The excellent agreement with measured data supports the employed strategy of using a magnetic configuration with $\mathfrak{M}$ closest to the measured normalized magnetization value.

We have tested the sensitivity of phonons to independent changes of magnetization, volume, and temperature. We find that (i) holding $T$ and $V$ fixed, increased magnetic disorder produces an overall softening of phonon modes, and (ii) holding $T$ and $\mathfrak{M}$ fixed, decreasing $V$ stiffens phonon modes. Interestingly, upon performing calculations at $T$ = 0$K$, 50$K$, 300$K$, and 600$K$, but keeping the magnetic configuration and volume fixed, the phonon dispersions were found to be nearly independent of $T$. Thus, the phonons are insensitive to larger atomic displacements produced by increasing temperature when $V$ and M are held fixed. This points to possible weak anharmonicity in the Invar bonding since the larger displacement amplitudes with increasing $T$ cause the constituent atoms to sample more anharmonic parts of the chemical bonds, but in Invar, this does not in itself significantly renormalize the phonon modes. In contrast, we have found that for bcc

Fe, the increase in lattice thermal atomic displacement amplitudes produced significant renormalization of the phonon modes [1].

Figure 3 investigates the effects of changing lattice thermal disorder ("lattice temperature") and magnetic thermal disorder ("magnetic temperature") independently. It compares the calculated pressures, $P_\mathfrak{M}$, for Invar (left panel) and bcc Fe (right panel) for two different lattice temperatures. For Invar, these are 50K and 296K, while for bcc Fe they are 300K and 600K. These temperatures reflect low and medium values relative to the respective transition temperatures of around 500K and 1043K. The lattice constants for Invar (3.5845Å) and Fe (2.83Å) are chosen to give roughly zero pressure at the lower of the two temperature values for the points matching the measured magnetizations at those temperatures. These lattice constant values are in good agreement with measured values [27, 32]. The figure demonstrates the relative effects of lattice and magnetic thermal disorder contributions to pressure in Invar compared with those for bcc Fe. A positive contribution to pressure works to expand the crystal (increase the lattice constant) while a negative contribution to pressure works to contract the lattice (decrease lattice constant). From this plot, we can make several conclusions. First, it is apparent that, in both invar and bcc Fe, increasing magnetic disorder (decreasing $\mathfrak{M}$) lowers the pressure i.e. the magnetic thermal disorder acts to reduce the lattice constant corresponding to a negative contribution to thermal expansion. Second, the magnitude of this effect is significantly larger in Invar than it is in bcc Fe. Specifically, comparing the spreads of the pressure ranges between fully-aligned (yellow) to random paramagnetic (blue) shows that this range is about 6 times larger in Invar than it is in bcc Fe. Thus, the same increase in magnetic thermal disorder produces a larger negative contribution to the thermal expansion in Invar than is the case for bcc Fe. Third, note the relative shift of each vertical set of points between the two temperatures for each material. Comparing points of equivalent magnetization at the two temperatures represents the effect of the lattice thermal disorder on pressure: an increase in pressure with increasing temperature corresponds to positive thermal expansion (i.e. the lattice wants to expand to reduce the pressure back to zero). The smaller pressure differences evident in Invar show that the contribution to its thermal expansion from increasing thermal lattice disorder is suppressed compared with bcc Fe.

Figure 4 plots the $P_\mathfrak{M}$ values for Invar (left panel) and bcc Fe (right panel) for a subset of the magnetizations shown in Fig. 3. The point marked "a" in the left panel corresponds to the zero pressure point for Invar at 50K, obtained for the lattice constant, 3.5845Å. The point marked "c" is calculated using the same 50K lattice constant, but now using thermal lattice configurations for 296K and the measured room temperature magnetization value of $\mathfrak{M} = 0.8$. The small value of $P$ found at point c (296K) is consistent with a near-zero thermal expansion over the 50K-300K range [33].

The lack of thermal expansion is found to be a consequence of a cancellation between the positive contribution from the lattice thermal disorder and the negative contribution from that of magnetism, as has been postulated before. This can be seen as follows. Starting at $P(T = 50K)$ (point a) and increasing $T$ to 296K while keeping the normalized magnetization, $\mathfrak{M}$, fixed at the 50K value brings us to point b. In this move from a to b, the only change is that lattice thermal disorder has increased. Thus, the increase seen in $P_\mathfrak{M}$ upon moving from a to b is caused by the increase in thermal lattice disorder. But in reality, $\mathfrak{M}$ also decreases when $T$ is increased, as the magnetic moments become more thermally disordered. Including this disorder by reducing $\mathfrak{M}$ to match the measured value at 296K reduces the pressure to the near-zero value at point c. Thus, for Invar, the negative contribution to thermal expansion from the increased thermal magnetic disorder is found to almost exactly cancel the positive contribution from the increased thermal lattice disorder. We emphasize that this remarkable result is captured using finite-temperature *ab initio* calculations within the DFT framework, and it demonstrates the extraordinary interplay between lattice and magnetic subsystems.

The results of similar calculations for bcc Fe are presented in the right panel of Fig. 4. In contrast to the findings in Invar, bcc Fe shows a large positive thermal expansion: point c is at a much higher pressure than point a, indicating that the zero-pressure volume for $T$=600K is noticeably larger than that for $T$=300K. This large increase is consistent with the measured large thermal expansion coefficient in bcc Fe [34]. It is dominated by the contribution from lattice thermal disorder, which can be seen by moving from point a to point b (increasing $T$ at fixed $V$ and $\mathfrak{M}$), which causes only a slightly larger increase in pressure compared with the move from a to c. Thus, we conclude that the contribution to thermal expansion from lattice thermal disorder is much larger in bcc Fe than in Invar. Furthermore, since the move from b to c in bcc Fe is quite small, we conclude that the negative contribution to thermal expansion in Fe from increasing magnetic disorder is negligible in bcc Fe in this temperature region. In fact, even increasing magnetic disorder in Fe to the fully-random paramagnetic extreme (point d) would not reduce the pressure enough to make the bcc result consistent with zero thermal expansion.

## V. Conclusions

In summary, a hybrid first principles theoretical framework for calculating temperature dependent phonons and examining thermal properties such as thermal expansion in magnetic materials of unknown entropy was presented. The approach integrates experimental measurements of the normalized magnetization into calculations of thermodynamic pressure in terms of configurational averaging over spin-lattice microstates. Thermal disorder in lattice and magnetic subsystems as well as spin-lattice coupling effects are naturally captured. The approach was tested on the classical Invar alloy $Fe_{0.65}Ni_{0.35}$ and on bcc Fe.

Excellent agreement with the measured room temperature phonon dispersions of Invar was achieved which, to our knowledge, has not been demonstrated previously. The results give a picture of the Invar effect in which (i) a positive contribution to thermal expansion from thermal lattice disorder is relatively small, possibly because of weak bond anharmonicity; (ii) contribution to thermal expansion from magnetic thermal disorder is relatively large, evidenced by the increased sensitivity of pressure to magnetic thermal disorder. The combination results in a near perfect cancellation between contributions from lattice and magnetic disorder in Invar. The behavior in bcc Fe is found to be in striking contrast, where a large positive lattice disorder contribution to thermal expansion is hardly reduced by the negative contribution from magnetic disorder. These findings are qualitatively consistent with prior predictions and measurements.

We emphasize that the results presented for thermal expansion in Invar and Fe are somewhat qualitative, and we do not claim to present a deep theoretical explanation for the anomalously small thermal expansion in Invar. Nevertheless, the proper trends that are identified along with the excellent agreement of the calculated phonon dispersions of Invar compared to experiment suggest that the new method may be a fruitful path for *ab initio* study of the thermal properties of magnetic materials of unknown entropy at finite temperature, which deserves further investigation.

**Acknowledgements:** This work was supported by the U.S. Department of Energy, Office of Science, Basic Energy Sciences, under Award # DE-SC0021071. M. H. and D. B. also acknowledge computational support from the Boston College Linux Cluster.

**Appendix A: Connecting configurational averaging to measured magnetization**

The microstates i.e. spin-lattice snapshots, enumerated by index $i$, are decomposed into $i = (\mathcal{L}, \mathcal{M})$, where $\mathcal{L}$ labels each lattice configuration, i.e. snapshot of atomic displacements, and $\mathcal{M}$ labels each magnetic configuration i.e. snapshot of magnetic moment orientations. Then, the configurational averaging to determine the DFT pressure, $\langle P_{DFT}[i] \rangle_{V,T}$, from Eq. 4, can be written:

$$\langle P_{DFT}[i] \rangle_{V,T} = \sum_{\mathcal{M}} \sum_{\mathcal{L}} w_{\mathcal{M}} w_{\mathcal{L}} P_{DFT}[(\mathcal{M}, \mathcal{L})] \tag{A1}$$

where $w_{\mathcal{M}}$ and $w_{\mathcal{L}}$ are the statistical weights of each magnetic and lattice state, respectively. By construction, each generated lattice configuration is an equally probable one for each $(V, T)$ so $w_{\mathcal{L}} = 1/N_{\mathcal{L}}$, where $N_{\mathcal{L}}$ is the total number of lattice configurations considered.

To identify the appropriate set of magnetic microstates, we start by defining a "pressure" for a given magnetic configuration, $\mathcal{M}$, as $P_{\mathcal{M}}$, such that

$$P_{\mathcal{M}}(V,T) = \frac{1}{2} P_{vib}(V,T) + \frac{1}{N_{\mathcal{L}}} \sum_{\mathcal{L}} P_{DFT}[(\mathcal{M}, \mathcal{L})] \tag{A2}$$

Then, the total pressure is:

$$P(V,T) = \sum_{\mathcal{M}} w_{\mathcal{M}} P_{\mathcal{M}}(V,T) \tag{A3}$$

To gain insight about how to group the magnetic states, the supercell atoms were placed at their equilibrium positions. Ideal lattice positions were used here rather than performing a magnetic relaxation for each thermally-displaced atomic configuration at each volume and temperature due to the high computational cost of such calculations. Then, various paramagnetic (fully random moment orientations) starting moment configurations were generated, and these were allowed to energetically relax towards the ground state, which was found to be ferromagnetic for both Invar and Fe. From the intermediate states visited during these magnetic relaxations, we extract a subset composing the set $\{\mathcal{M}\}$. Each $\mathcal{M}$ is characterized according to their normalized magnetizations, $\mathfrak{M}$, and their energy above the ground state. Fig. A1 plots the dependency of $P_{\mathcal{M}}$ (for a representative subset of $\mathcal{M}$) on amount of magnetic order, $\mathfrak{M}$, and energy above the ground state. From this figure can be seen the important conclusion that $P_\mathcal{M}$ is determined exclusively by $\mathfrak{M}$ and is virtually independent of energy. This is seen by the fact that the color is essentially determined by the vertical position of a point on the plot and is unaffected by horizontal location. Analogous behavior was observed for $T = 296K$. Fig. A2 plots the same data (with the addition of $T = 296K$ data) without any designation of energy. The conclusion from Fig. A1 can also be seen in this figure in that the vertical spread (different energies) of points with similar $\mathfrak{M}$ is small. This demonstrates that, for the purpose of calculating pressure, the relevant characteristic of magnetic disorder for a microstate is $\mathfrak{M}$, the amount of orientational disorder in the magnetic moments. Furthermore, Fig. A2 demonstrates that, for the $\mathfrak{M}$ values of interest in this study, $P_\mathcal{M}$ is a simple *linear* function of $\mathfrak{M}$.

**Figures**

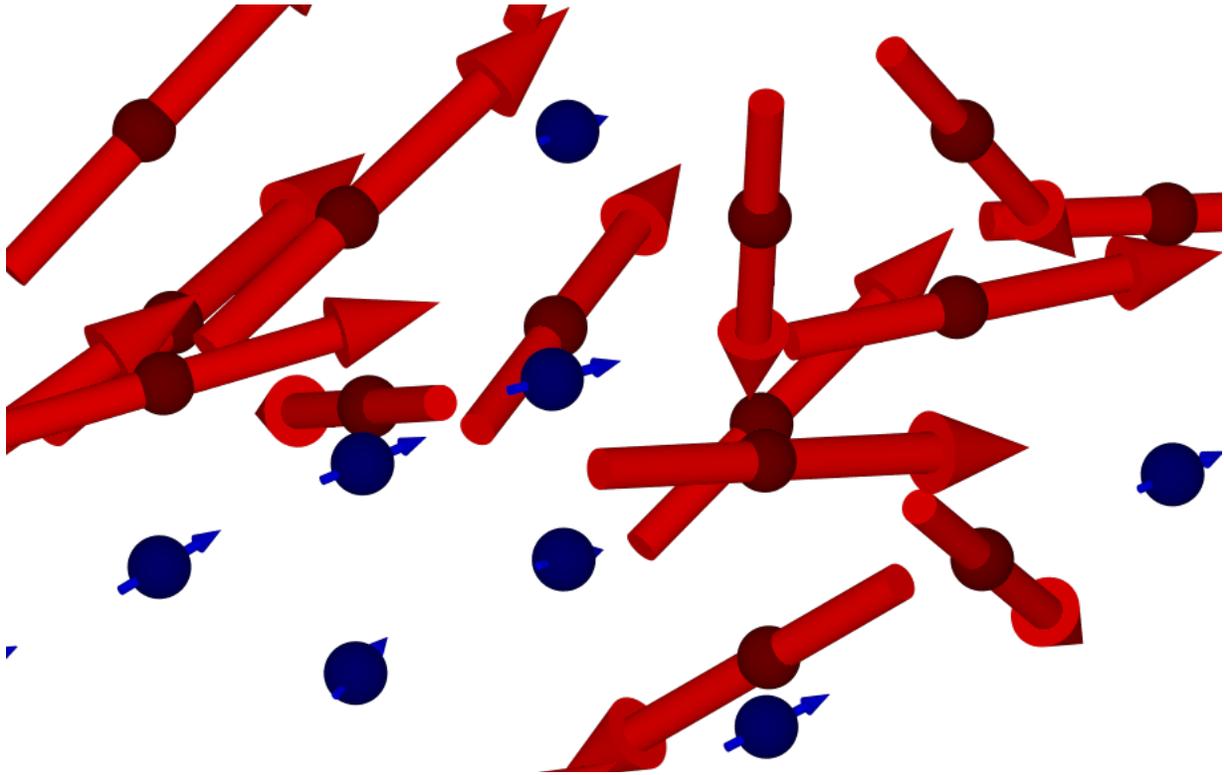

Figure 1. Example of a thermal spin-lattice snapshot (microstate) for the Invar calculation. Circles represent Fe (red) and Ni (blue) atoms while arrows represent corresponding magnetic moments. Note that atoms are thermally displaced from their ideal lattice positions and moments have disordered orientations.

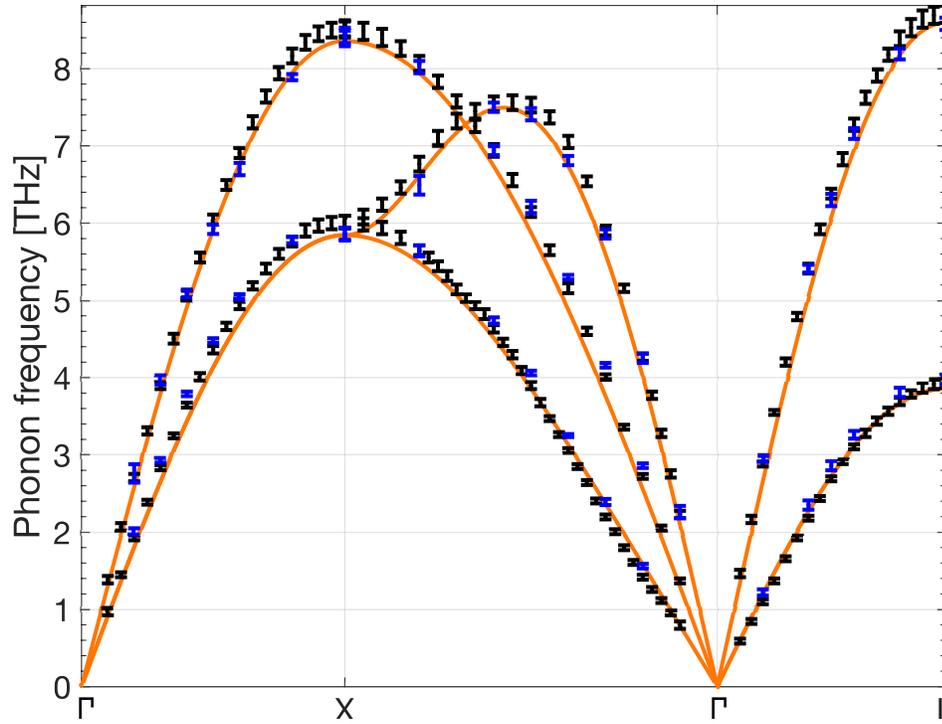

Figure 2: Calculated phonon dispersions for the classical Invar alloy at room temperature compared to measurement. Blue error bars are measurements of $Fe_{0.7}Ni_{0.3}$ from Ref. 25 while black error bars are measured data for $Fe_{0.65}Ni0.35$ from Ref. 26.

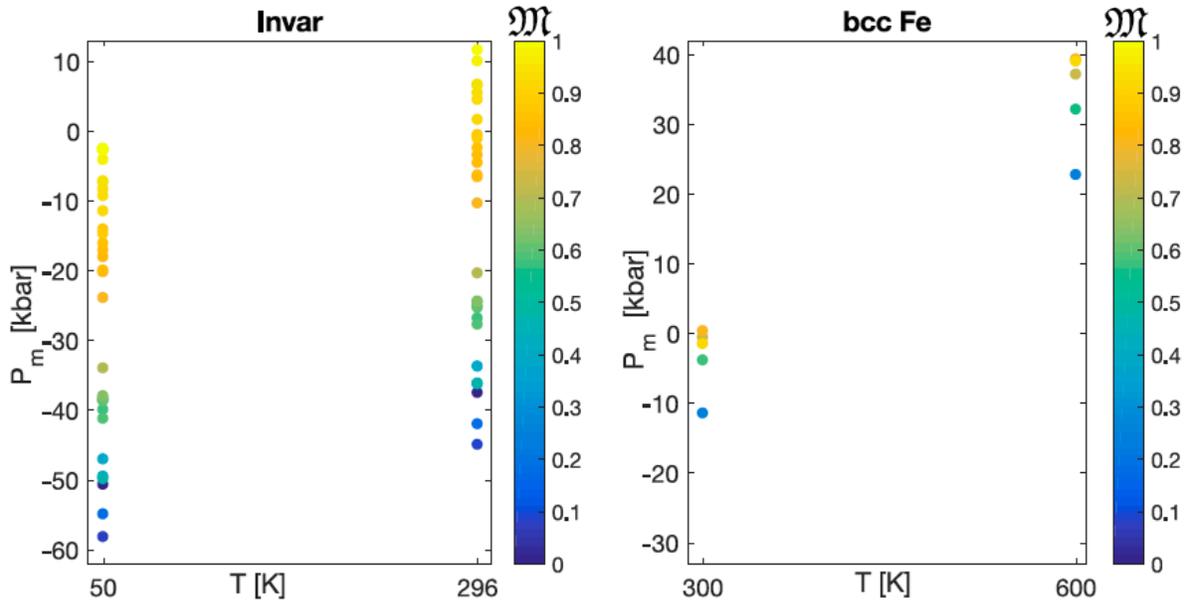

**Figure 3:** Left panel: Calculated $P_\mathfrak{M}$ for Invar at ambient volume for $T$=50K and normalized magnetization, $\mathfrak{M}$, ranging from ferromagnetic, $\mathfrak{M}$=1, to paramagnetic, $\mathfrak{M}$=0 (see colorbar for normalized magnetization scale). Calculated $P_\mathfrak{M}$ values at $T$=296K use the ambient volume at 50K. Right panel: Same for bcc Fe for $T$=300K and $T$=600K.

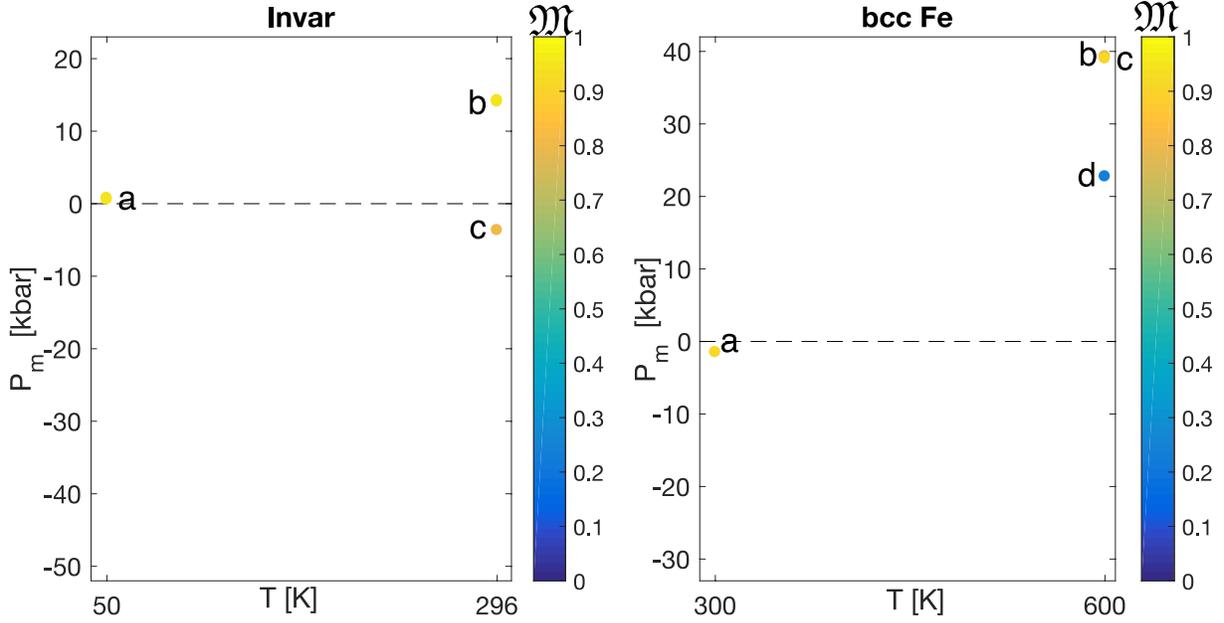

**Figure 4**: Calculated $P_\mathfrak{M}$ a subset of values in Fig. 3, for Invar (left panel) and for bcc Fe (right panel). See Fig. 3 caption for axes/colormap definitions. Labeled points for Invar are: point a: calculated $P_\mathfrak{M}(T = 50K)$ using $\mathfrak{M}$ consistent with measured magnetization at 50K; point b: $P_\mathfrak{M}(T = 296K)$ for same $\mathfrak{M}$ as at 50K. point c: calculated $P_\mathfrak{M}(T = 296K)$ with $\mathfrak{M}$ consistent with measurement at 296K; Labeled points a, b, c for bcc Fe have same meaning as corresponding points for Invar but at 300K and 600K. Point d: $P_\mathfrak{M}$ for a paramagnetic configuration.

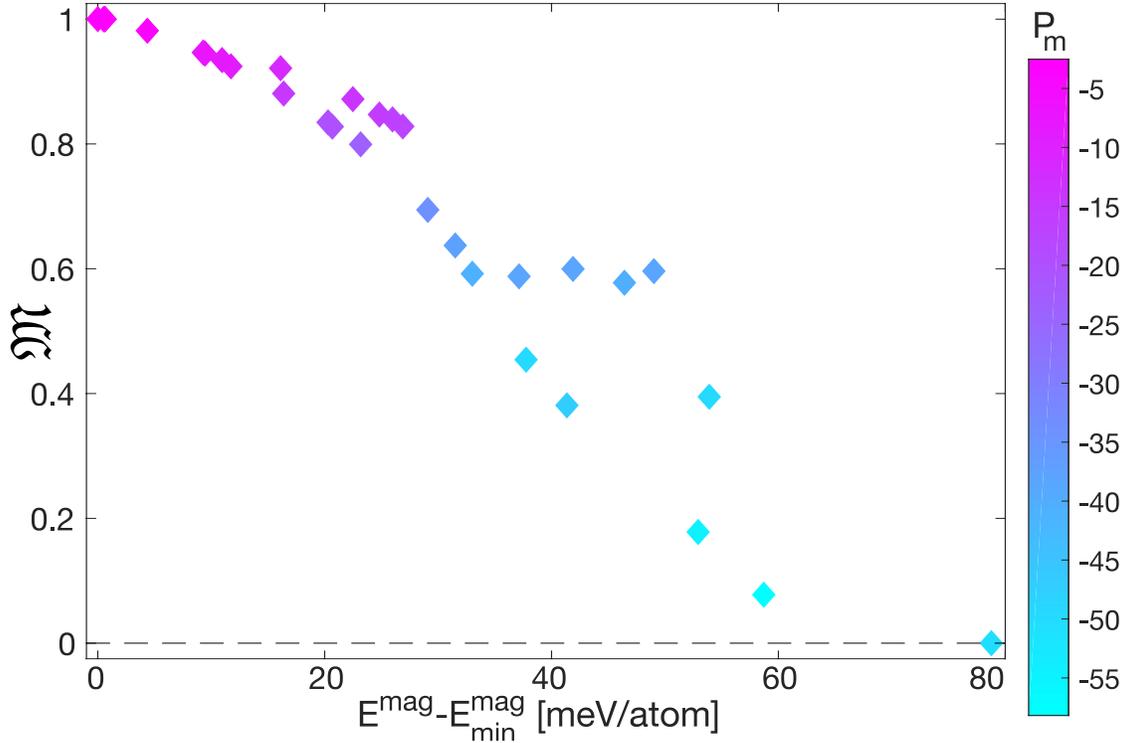

Figure A1: Dependence of $P_\mathcal{M}$ for Invar upon degree of disorder, $\mathfrak{M}$, and energy of the magnetic state. Values of $P_\mathcal{M}$ are given by color (see colorbar for values in kbar). $E^{mag}$ is the energy of the given magnetic configuration with atoms in the ideal lattice positions; $E^{mag}_{min}$ is the lowest energy of the plotted configurations. All points correspond to a lattice temperature of 50K.

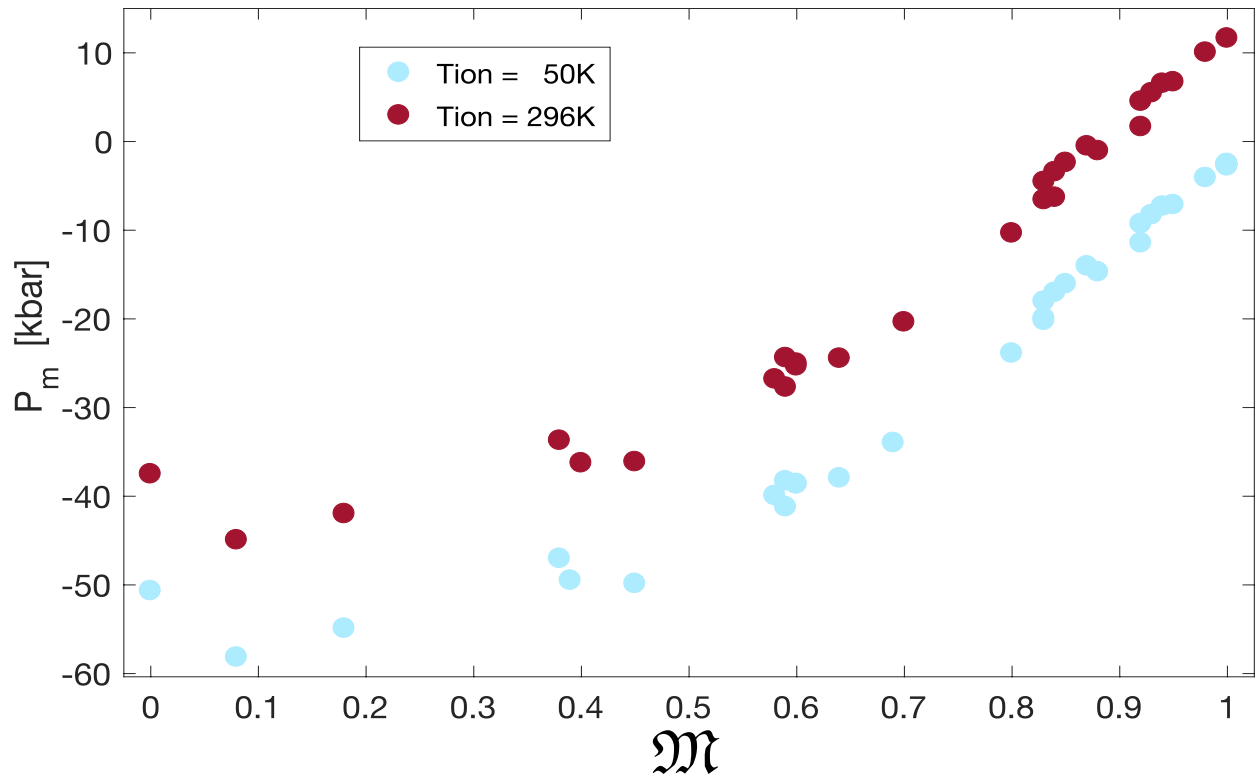

Figure A2: Dependency of $P_\mathfrak{M}$ for Invar on degree of magnetic disorder, $\mathfrak{M}$ for lattice temperatures of 50K and 296K. The T=50K data is the same as in Fig. A1. For the $\mathfrak{M}$ range of interest, $P_\mathfrak{M}$ is a well-determined linear function of $\mathfrak{M}$.